\documentclass[a4paper,12pt]{article}
\usepackage{graphicx,amssymb,bm,latexsym}
\newcommand{\bea}{\begin{eqnarray}}
\newcommand{\ena}{\end{eqnarray}}
\newcommand{\vs}[1]{\vspace{#1 mm}}

\renewcommand{\a}{\alpha}

\newcommand{\p}[1]{(\ref{#1})}

\begin{document}

\begin{titlepage}

\begin{flushright}
KU-TP 032
\end{flushright}

\vs{10}
\begin{center}
{\Large\bf  Thermodynamics of Black Holes in Ho\v{r}ava-Lifshitz Gravity}
\vs{15}

{\large Rong-Gen Cai$^{a,}$\footnote{e-mail address: cairg@itp.ac.cn},
Li-Ming Cao$^{b,}$\footnote{e-mail address: caolm@apctp.org},
Nobuyoshi Ohta$^{c,}$\footnote{e-mail address: ohtan@phys.kindai.ac.jp}}
\vs{15}

{\em $^a$ Key Laboratory of Frontiers in Theoretical
Physics, Institute of Theoretical Physics, Chinese Academy of Sciences,
P.O. Box 2735, Beijing 100190, China \\
Kavli Institute for Theoretical Physics China (KITPC),
Chinese Academy of Sciences,
P.O. Box 2735, Beijing 100190, China \\
$^{b}$Asia Pacific Center for Theoretical Physics, Pohang,
Gyeongbuk 790-784, Korea \\
$^{c}$Department of Physics, Kinki University, Higashi-Osaka, Osaka
577-8502, Japan }
\end{center}

\vspace{10mm}
\centerline{{\bf{Abstract}}} \vspace{5mm}

By using the canonical Hamiltonian method, we obtain the mass and
entropy of the black holes with general dynamical coupling constant
$\lambda$ in Ho\v{r}ava-Lifshitz Gravity. Regardless of whether the
horizon is sphere, plane or hyperboloid, we find these black holes
are thermodynamically stable in some parameter space and unstable
phase also exists in other parameter space. The relation between the
entropy and horizon area of the black holes has an additional
coefficient depending on the coupling constant $\lambda$, compared
to the $\lambda=1$ case. For $\lambda=1$, the well-known coefficient
of one quarter is recovered in the infrared region.

\end{titlepage}
\section{Introduction}
Recently a field theory model for a UV complete theory of gravity
was proposed by Ho\v{r}ava~\cite{Horava}, which is a
non-relativistic renormalisable theory of gravity and reduces to
Einstein's general relativity at large scales for the dynamical
coupling constant $\lambda=1$. Much attention has been paid to this
gravity theory~\cite{Visser}--\cite{Kehagias:2009is}. The authors of
\cite{LMP} found some static spherically symmetric black hole
solutions in Ho\v{r}ava-Lifshitz theory and \cite{CCO} presented
topological black hole solutions and discussed the associated
thermodynamic properties with those black hole solutions.

In the $(3+1)$-dimensional ADM formalism, where the metric can be
written as
\begin{equation}
\label{eq1} ds^2=-N^2 dt^2 +g_{ij}(dx^i-N^idt)(dx^j-N^jdt),
\end{equation}
and for a spacelike hypersurface with a fixed time, its extrinsic
curvature $K_{ij}$ is
\begin{equation}
\label{eq2} K_{ij}=\frac{1}{2N}(\dot
g_{ij}-\nabla_iN_j-\nabla_jN_i),
\end{equation}
where a dot denotes a derivative with respect to $t$ and covariant
derivatives defined with respect to the spatial metric $g_{ij}$.
The action of Ho\v{r}ava-Lifshitz theory is~\cite{Horava}
\begin{eqnarray}
\label{eq3}
I &=& \int dtd^3x ({\cal L}_0 +{\cal L}_1), \\
 {\cal L}_0 &=& \sqrt{g}N \left \{\frac{2}{\kappa^2}
(K_{ij}K^{ij}-\lambda K^2) +\frac{\kappa^2\mu^2 (\Lambda
R-3\Lambda^2)}{8(1-3\lambda)}\right \},  \nonumber \\
 {\cal L}_1  &=& \sqrt{g}N \left \{\frac{\kappa^2\mu^2(1-4\lambda)}{32(1-3\lambda)}R^2
-\frac{\kappa^2}{2\omega^4}Z_{ij} Z^{ij} \right\}\, ,\nonumber
\end{eqnarray}
where
\begin{equation}
Z_{ij}=C_{ij}-\frac{\mu\omega^2}{2}R_{ij}\, .
\end{equation}
and $\kappa^2$, $\lambda$, $\mu$, $\omega$ and $\Lambda$ are
constant parameters and the Cotten tensor, $C_{ij}$, is defined by
\begin{equation}
\label{eq4} C^{ij}=\epsilon^{ikl} \nabla_k \left (R^j_{\
l}-\frac{1}{4}R\delta^j_l\right) = \epsilon^{ikl}\nabla_k R^j_{\ l}
-\frac{1}{4}\epsilon^{ikj}\partial_kR.
\end{equation}
In (\ref{eq3}), the first two terms are the kinetic terms, while the
others give the potential of the theory in the so-called
``detailed-balance" form.

Comparing the action to that of general relativity, one can see that
the speed of light, Newton's constant and the cosmological constant
are
\begin{equation}
\label{eq5}
c=\frac{\kappa^2\mu}{4}\sqrt{\frac{\Lambda}{1-3\lambda}}, \ \
G=\frac{\kappa^2 c}{32\pi}, \ \ \tilde \Lambda=\frac{3}{2}\Lambda,
\end{equation}
respectively. Let us notice that when $\lambda=1$, the first three
terms in (\ref{eq3}) could be reduced to the usual ones of
Einstein's general relativity. However, in Ho\v{r}ava-Lifshitz
theory, $\lambda$ is a dynamical coupling constant, susceptible to
quantum correction~\cite{Horava}. In addition, we see from
(\ref{eq5}) that when $\lambda >1/3$, the cosmological constant
$\Lambda$ must be negative. However, the cosmological constant can
be positive if we make an analytic continuation $\mu \to i\mu, w^2
\to -iw^2$~\cite{LMP}. In this paper, we consider the former case
with a negative cosmological constant.

The equations of motion for the action~\p{eq3} are given  as
\cite{Kiritsis,LMP}
\begin{equation}
\label{EOMLapse} \frac{2}{\kappa^2}\left(K_{ij}K^{ij}-\lambda
K^2\right)-\frac{\kappa^2 \mu^2 (\Lambda
R-3\Lambda^2)}{8(1-3\lambda)}-\frac{\kappa^2 \mu^2
(1-4\Lambda)}{32(1-3\lambda)}R^2+\frac{\kappa^2}{2 \omega^4
}Z_{ij}Z^{ij}=0\, ,
\end{equation}
\begin{equation}
\label{EOMShift} \nabla_k\left(K^{kl}-\lambda Kg^{kl}\right)=0\, ,
\end{equation}
and
\begin{equation}
\label{EOMgij}
\frac{2}{\kappa^2}E^{(1)}_{ij}-\frac{2\lambda}{\kappa^2}E^{(2)}_{ij}+\frac{\kappa^2
\mu^2 \Lambda}{8(1-3\lambda)}E^{(3)}_{ij}+\frac{\kappa^2 \mu^2
(1-4\lambda)}{32(1-3\lambda)}E^{(4)}_{ij}-\frac{\mu\kappa^2}{4\omega^2}E^{(5)}_{ij}
-\frac{\kappa^2}{2\omega^4}E^{(6)}_{ij}=0
\end{equation}
where the tensors $E^{(1)}_{ij}$, $E^{(2)}_{ij}$, $E^{(3)}_{ij}$,
$E^{(4)}_{ij}$, $E^{(5)}_{ij}$ and $E^{(6)}_{ij}$ are composed by
$K_{ij}$, $g_{ij}$, $N$, $N_i$ and their covariant derivatives with
respect to the three dimensional metric. The explicit forms of these
tensors can be found, for example, in Ref.~\cite{LMP}.

The static, spherically symmetric solutions have been found
in~\cite{LMP}. The solutions for $\lambda=1$ are asymptotically
$AdS_4$ and may be of some interest in AdS/CFT correspondence. The
solutions have been extended to general topological black holes, in
which the two dimensional sphere as black hole horizon has been
generalized to two dimensional constant curvature spaces, and their
thermodynamic properties including the definition of the mass and
entropy are discussed in Ref.~\cite{CCO} for the case of
$\lambda=1$. Another remarkable point is that the solution of
general relativity is found not always to be recovered at large
distance~\cite{LMP}. For large distance, the Einstein theory could
only arise for the case with $\lambda =1$. This indicates that the
infrared region of Ho\v{r}ava-Lifshitz theory can deviate from
Einstein's general relativity.

It is extremely interesting to study the properties of this kind of
black holes for general $\lambda$. This is because all of the
important properties of the Ho\v{r}ava-Lifshitz theory may not be
revealed by just studying some special cases like $\lambda=1$. It is
possible that some important properties of the theory will emerge in
the case with general $\lambda$. Black hole thermodynamics can give
some lights on some aspects of the quantum effects of gravity.
Compared with other UV complete theories, such as string theory,
Ho\v{r}ava-Lifshitz theory has quite different UV behavior. So it is
an urgent problem to examine the effect of quantum gravity by
studying the thermodynamics of black holes in this theory. Even at
the semi-classical level, it is also interesting to study the
thermodynamical stability of these black holes. In this paper, we
extend our previous study~\cite{CCO} of black hole thermodynamics of
topological black holes for $\lambda=1$ to general values of
$\lambda$. In particular we discuss how to define the mass and
entropy in this general situation by using the canonical Hamilton
formulation~\cite{BTZ, CS, RT, MTZ}. In \cite{CCO} we have used the
first law of black hole thermodynamics to find the entropy
expression of the topological black hole solutions in
Ho\v{r}ava-Lifshitz theory. Both methods give the same result.

In the course of writing this paper, a paper appeared~\cite{MK}
which discusses related subject, but the proper definitions of the
mass and entropy are not discussed.

\section {Topological Black Holes with General $\lambda$}

Here we briefly review the topological black hole solutions with
general $\lambda$ which have been presented  in~\cite{CCO}. Assume
that the metric of the black hole is given by
\begin{equation}
ds^2=-\tilde{N}^2 f(r)dt^2+\frac{dr^2}{f(r)}+r^2 d\Omega_k^2\, .
\label{action}
\end{equation}
In terms of the new function $F$ defined by \bea F(r)=k-\Lambda
r^2-f(r), \label{3eq1} \ena the action takes the form \bea I =
\frac{\kappa^2\mu^2 \Omega_k}{8(1-3\lambda)}\int dt dr \tilde N
\left \{ \frac{(\lambda-1)}{2} F'^2
 - \frac{2\lambda}{r} FF' +\frac{(2\lambda-1)}{r^2} F^2 \right \}.
\label{3eq2}
\ena
After variation this reduced action, we get the
equations of motion
\bea
\label{3eq31}
&& 0 = \left(\frac{2\lambda}{r}F-(\lambda-1)F' \right) \tilde N'
+(\lambda-1)\left(\frac{2}{r^2}F-F'' \right)\tilde N, \\
&& 0= (\lambda-1)r^2 F'^2 -4\lambda rFF'+2(2\lambda-1)F^2.
\label{3eq32} \ena The latter is easily solved to give \bea F(r)= \a
r^{s}, \label{3eq4}, \ena where \bea s =
\frac{2\lambda\pm\sqrt{2(3\lambda-1)}}{\lambda-1}, \label{s} \ena
and then the first gives \bea \tilde N = \gamma r^{1-2s} ,
\label{3eq5} \ena where $\a$ and $\gamma$ are both integration
constants. If we use the usual units in gravity theory, $\gamma$ has
an inverse dimension of $r^{1-2s}$. For $\lambda=1$ or $s=1/2$ case,
in which it is dimensionless, one can set $\gamma=1$ by rescaling
the time coordinate $t$~\cite{CCO}.
When $\a=0$ or $F=0$, Eq.~\p{3eq31} does not
restrict $\tilde N$. For the case $k=1$, our solution reduces to the
one given in Ref.~\cite{LMP}. Substituting this metric into the
equations (\ref{EOMLapse}), (\ref{EOMShift}) and (\ref{EOMgij}), we
find that this metric with $f$ and $\tilde{N}$ above indeed
satisfies the equations of motion.

It is interesting to note that there are two branches in (\ref{s}).
It is easy to find that the range of $s$ is $(-1,2)$ for the
negative branch in the case of $\lambda >1/3$, which will be assumed
in the present paper. Note that the exponent $s$ of Eq.~\p{3eq4} for
the negative branch is always less than $2$ for positive $\lambda$,
and thus the $r^2$ term in the metric function~\p{3eq1} dominates in
large distance. This suggests that negative branch solution has some
asymptotic behavior of AdS spacetime. On the other hand,  the
positive branch $s$ gives a power larger than $2$ for $\lambda>1$.
In that case, the $F$ term will dominate at large distances and the
solution will have a cosmological horizon-like if $\alpha>0$. In
this case, the physical meaning of the solution is not very clear
since the solution is not asymptotic to the vacuum solution with
$\alpha=0$ at infinity.  In the case of $1>\lambda>1/3$, the
positive branch gives a negative power. In this case, the solution
seemingly makes sense. However, some physical quantities are not
well defined in this case. Therefore in the present paper we limit
ourselves to the negative branch with $s$ in the range $s\in [-1, 2)$.

\section{Black Hole Thermodynamics}

In this section, we discuss black hole thermodynamics by using
the canonical Hamilton formulation~\cite{BTZ,CS, RT, MTZ, Peca:1998dv, Peca:1998cs}.
The partition function for a thermodynamical ensemble is identified with the
Euclidean path integral in the saddle point approximation around
Euclidean continuation of the classical solution.

For our solutions, their asymptotically behaviors are complicated.
Those solutions are neither asymptotically flat  nor asymptotically
AdS. As a result, the definitions of ADM mass or conformal
mass~\cite{Ashtekar:1984,Ashtekar:1999jx} are not applicable here
(in fact, one will obtain a divergent result if naively uses these
definitions). On the other hand, we find that the canonical
Hamiltonian method works well in our case and  enables us to define
finite mass associated with those solutions.

Consider the Euclidean continuation of the action of the topological
black holes for general $\lambda$ in Hamiltonian form ($I\rightarrow
-I_E$)
\begin{equation}
I_E=\int d^3xdt\left[\pi^{ij}\dot{g}_{ij} -
N\mathcal{H}-N^i\mathcal{H}_i\right]+B\, ,
\end{equation}
where the $B$ is a boundary term. $N$ and $N_i$ are lapse function
and shift function respectively. In our case, the lapse function is
given by $N^2=\tilde{N}^2 f$. Since we are considering static black
hole case, we need not give the explicit form of the momentum
$\mathcal{H}_i$ and conjugate $\pi^{ij}$ of $\dot{g}_{ij}$. For the
black hole metric~\p{action}, the Euclidean action is reduced to
\begin{equation}
I_E=-\beta~ \Omega_k\int_{r_+}^{\infty}\tilde{N}(r)\mathcal{H}(r)dr
+B\, ,
\end{equation}
where $\mathcal{H}(r)$ is given by
\begin{equation}
\mathcal{H}(r)=\frac{\kappa^2\mu^2 \Omega_k}{8(1-3\lambda)}\left \{
\frac{(\lambda-1)}{2} F'^2
 - \frac{2\lambda}{r} FF' +\frac{(2\lambda-1)}{r^2} F^2 \right \}\,
 ,
\end{equation}
$\beta$ is the period of Euclidean time and $r_+$ is the radius of
the black hole horizon defined by the largest root of $f(r)=0$ or
$F(r)=k-\Lambda r^2$ from \p{3eq1}. The Euclidean black
holes are static and satisfy the constraint $\mathcal{H}=0$. So the
Euclidean action is just the boundary term $B$. The existence of
this boundary term ensures that we can get correct equations of motion
from variation of the Euclidean action.

To avoid conical singularity at horizon of the Euclidean black hole
solution, we have to set the time period $\beta$ to
\begin{equation}
\label{Euclideanperiod} \beta (\tilde{N}(r) f'(r))|_{r=r_+}= 4\pi\,
,
\end{equation}
which gives the temperature of the black hole
\begin{equation}
T=\frac{1}{\beta}=\frac{\gamma}{4\pi r_+^{2s}}  \left[-\Lambda r_+^2
(2-s)- s k\right]\, .
\end{equation}
When $\lambda \rightarrow 1$, from L'Hospital rule, we have $s=1/2$.
In this case, we obtain
\begin{equation}
T=\frac{\gamma}{8\pi r_+}  \left[-3\Lambda r_+^2 -  k\right]\, ,
\end{equation}
which is just the temperature given in Ref.~\cite{CCO} for
$\gamma=1$.

In the canonical ensemble, this temperature should be kept fixed
under the variation of the action. From the variation of the
Euclidean action, we find that the variation of the boundary term is
given by
\begin{equation}
\label{variationB} \delta B= \delta B|_{\infty}-\delta
B|_{r_+}=-\beta \frac{\kappa^2\mu^2 \Omega_k}{8(1-3\lambda)}
\left[\lambda \left(\frac{2}{r}\tilde{N}F\delta F\right)-(\lambda-1)
(\tilde {N}F' \delta F)\right]_{r_+}^{\infty}\, .
\end{equation}
To get equations of motion, we need not  know the explicit form of
$\delta F$ or $\delta \tilde{N}$, but here we need them, which can
be obtained from our solutions~\p{3eq4}. Another point that should
be noted is that the coordinate $r$ is invariant under the
variation; this is the same as in the process of variation to get
the equations of motion (\ref{3eq31}) and (\ref{3eq32}). Near
infinity, from the expression of $F$ in \p{3eq4}, we find
\begin{equation}
\delta F=\delta \left(\alpha r^{s}\right)=r^{s}\delta \alpha\, ,
\end{equation}
so $\alpha$ is the only thermodynamic parameter. From the expression
$\tilde{N}$ in \p{3eq5}, we have
\begin{equation}
\label{variation1} \lambda \left.\left(\frac{2}{r}\tilde{N}F\delta
F\right)\right|_{\infty} = \frac{2\lambda}{r} \left(\gamma
r^{1-2s}\right)\left(\alpha r^{s}\right) \left(r^{s}\delta
\alpha\right) =2\lambda \gamma \alpha \delta \alpha\, .
\end{equation}
Although $\delta F$, $F$ or $\tilde{N}$ diverge at infinity,
the combination $(2/r)\tilde{N}F\delta F$ is finite. In fact, even
for the case $\lambda=1$ or $s=1/2$, the $\delta F$ is divergent as
$\sqrt{r}$, but $(2/r)F\delta F$ is finite. The only special point
of this case is that $\tilde{N}$ is constant.  Certainly, for the
case $\lambda=1/3$ or $s=-1$, this combination is also finite, and
$\delta F$ has good behavior like $1/r$ (but $\tilde{N}$ rapidly
increases asymptotically as $r^3$). For the case $1/2 < s <2$,
$\delta F$ increases faster than $s=1/2$ case. However, since
$(2/r) \tilde{N}$ decreases so as to cancel this divergence,
we can always get finite result. Similarly, we find
\begin{equation}
\label{variation2} (\lambda-1) \left.\left(\tilde {N}F' \delta
F\right)\right|_{\infty}= \left(2\lambda - \sqrt{6\lambda
-2}\right)\gamma\alpha \delta \alpha\, .
\end{equation}
Combining equations (\ref{variationB}), (\ref{variation1}) and
(\ref{variation2}), we get the variation of the boundary term at
infinity
\begin{equation}
\delta B|_{\infty} = -\beta \frac{\kappa^2\mu^2
\Omega_k}{8(1-3\lambda)} \sqrt{6\lambda -2}~\gamma\alpha \delta
\alpha\, .
\end{equation}
This suggests that this boundary term is given by
\begin{equation}
B|_{\infty}=\beta \frac{\sqrt{2}\kappa^2\mu^2
\Omega_k}{16\sqrt{3\lambda -1}} \gamma\alpha^2 \, .
\end{equation}
For the boundary at the horizon, the variation of $F$ is given
by~\cite{BTZ, MTZ}
\begin{equation}
\delta F|_{r_+}= \left(\frac{\partial F}{\partial
f}\right)_{r_+}\left[\delta f\right]_{ r_+}\, .
\end{equation}
Since on the horizon, we have
\begin{equation}
\left[\delta f\right]_{ r_+}+ \left(\frac{d
f}{dr}\right)_{r_+}\delta r_+=0\, ,
\end{equation}
then
\begin{equation}
\delta F|_{r_+}= -\left(\frac{\partial F}{\partial f}\right)_{r_+}
\left(\frac{d f}{dr}\right)_{r_+}\delta r_+=\left(\frac{d
f}{dr}\right)_{r_+}\delta r_+\, .
\end{equation}
As a result, by using the relation (\ref{Euclideanperiod}), we
arrive at
\begin{equation}
\delta B|_{r_+}= -\frac{\pi\kappa^2\mu^2 \Omega_k}{2(1-3\lambda)}
\left[ \frac{2\lambda}{r_+}F(r_+)-(\lambda -1)F'(r_+)\right]\delta
r_+\, .
\end{equation}
This way we obtain the black hole entropy
\begin{equation}
B|_{r_+}=S\, ,
\end{equation}
where
\begin{equation}
S=\frac{\pi\kappa^2\mu^2 \Omega_k}{\sqrt{2(3\lambda-1)}}\int
G(r_+)dr_+ + S_0\, ,
\end{equation}
with integration constant $S_0$. The integration constant $S_0$
should be fixed as discussed in \cite{CCO}. The integrand $G(r_+)$
is given by
\begin{equation}
G(r_+)= \frac{1}{r_+}F(r_+)=\frac{1}{r_+}(k-\Lambda r_+^2)\, .
\end{equation}
For the on-shell solution,  the Euclidean action is just the
boundary term. Namely, we have
\begin{equation}
I_E= B= B|_{\infty}- B|_{r_+}\, ,
\end{equation}
which gives
\begin{equation} I_E=\beta
\frac{\sqrt{2}\kappa^2\mu^2 \Omega_k}{16\sqrt{3\lambda -1}} \gamma
\alpha^2 - S\, .
\end{equation}
Since the Euclidean action has relation to free energy $F_e$ by \bea
I_E=\beta F_{e}=\beta  M-S, \ena where $S$ is the entropy and $M$ is
the mass. Thus we get the mass, temperature and the entropy of the
black holes as follows. \bea &&M=\frac{\sqrt{2}\kappa^2\mu^2
\Omega_k}{16\sqrt{3\lambda -1}}
\gamma \alpha^2 \, , \\
&&T=\frac{\gamma}{4\pi r_+^{2s}}  \left[-\Lambda r_+^2 (2-s)- s
k\right]\,
, \\
&& S=\frac{ \pi\kappa^2\mu^2 \Omega_k}{\sqrt{2(3\lambda-1)}}\left[k
\ln (\sqrt{-\Lambda }r_+) + \frac{1}{2}(\sqrt{-\Lambda}r_+)^2\right]
+ S_0\, . \ena We can also express the mass by radius of the horizon
\begin{equation}
M=\frac{\sqrt{2}\kappa^2\mu^2 \gamma \Omega_k}{16\sqrt{3\lambda
-1}}\frac{(k-\Lambda r_+^2)^2}{r_+^{2s}}\, .
\end{equation}
Defining $\ell^2=-1/\Lambda$ and using (\ref{eq5}) and (\ref{s}), we
have
\begin{equation}
M=\frac{c^3 }{16 \pi G}\left(\frac{1+s}{2-s}\right)\left(\gamma
\Omega_k\ell^{2-2s}\right) \left[\frac{k+(
r_+/\ell)^2}{(r_+/\ell)^{s}}\right]^2\, ,
\end{equation}
where $c$ is the light velocity defined in Eq.~(\ref{eq5}) and can
be re-expressed in terms of $s$ and $\ell$ instead of $\lambda$ and
$\Lambda$:
\begin{equation}
\label{lightvelocity}
c=\left(\frac{2-s}{1+s}\right)\left(\frac{\kappa^2
\mu}{4\sqrt{2}\ell}\right)\, .
\end{equation}
The temperature is given by
\begin{equation}
T=\frac{\gamma}{4\pi r_+^{2s}}\left[(r_+/\ell)^2 (2-s)-ks\right]\, .
\end{equation}
The entropy can also be expressed as
\begin{equation}
S=\frac{c^3 }{4
G}\left(\frac{1+s}{2-s}\right)\left(\Omega_k\ell^{2}\right)\left[k\ln
\left(\frac{r_+}{\ell}\right)^2
+\left(\frac{r_+}{\ell}\right)^2\right]+S_0\, .
\end{equation}
When $s=1/2$ or $\lambda =1$, It goes to the one obtained in
\cite{CCO}. It is easy to confirm that these thermodynamical
quantities satisfy the first law of thermodynamics
\begin{equation}
dM=TdS\, .
\end{equation}
Note that in Ref.~\cite{CCO}, we have derived the entropy using the first law,
but here we have shown that the canonical Hamiltonian formalism allows us
to define the entropy which satisfies the first law.



In general we cannot determine whether the black holes are
thermodynamic stable or not since one cannot fix the integration
constant $S_0$ here. As argued in \cite{CCO}, to fix the integration
constant $S_0$, one has to invoke the quantum theory of the gravity.
For the Ricci flat black holes with $k=0$, the logarithmic term
is absent and the entropy is proportional to the horizon area.
In this case we can set $S_0 =0$ by the assumption that black hole entropy
vanishes when horizon goes to zero. Thus we have the free energy of
the black hole
\begin{equation}
F_{e}\sim \gamma (s-1)M.
\end{equation}
This result is also valid for large black holes.
We see that the large black holes and those with $k=0$ are
always thermodynamically stable globally when $s\le 1$.  When $s>1$,
the free energy turns out to be positive, which means that
the black hole is thermodynamically unstable globally. This is quite
different from the situation in Einstein's general relativity, where large
AdS  black holes are always thermodynamically stable regardless of the
horizon topology.

\section{Some Special Cases}

\subsection{Einstein gravity: $\lambda\rightarrow 1$}

For the case $\lambda=1$, we have $s=1/2$.  Those thermodynamic
quantities become
\begin{equation}
M=\frac{c^3}{16\pi G}\left(\Omega_k \ell\right) \left[\frac{k+(
r_+/\ell)^2}{(r_+/\ell)^{1/2}}\right]^2\, ,
\end{equation}
\begin{equation}
T=\frac{1}{8\pi r_+}\left[3(r_+/\ell)^2 -k\right]\, ,
\end{equation}
and
\begin{equation}
S=\frac{c^3}{4G}\left(\Omega_k\ell^2\right) \left[k\ln
\left(\frac{r_+}{\ell}\right)^2
+\left(\frac{r_+}{\ell}\right)^2\right]+S_0\, .
\end{equation}
Here we have set $\gamma = 1$ by the rescaling of the time. These
are just what we have found in~\cite{CCO}. Since this case has been
discussed in some details in~\cite{CCO}, we will not repeat the
discussions here.

\subsection{Black holes with flat horizon}

In this case, the thermodynamic quantities have the forms
\bea
&& M= \frac{c^3 }{16 \pi
G}\left(\frac{1+s}{2-s}\right)\left(\gamma\Omega_k\ell^{2-2s}\right)(
r_+/\ell)^{2(2-s)}\, , \\
&& T=\frac{\gamma}{4\pi }\ell^{-2s}(2-s)(r_+/\ell)^{2 -2s}\, ,\\
&& S=\frac{c^3
}{4G}\left(\frac{1+s}{2-s}\right)\left(\Omega_k\ell^{2}\right)
\left(\frac{r_+}{\ell}\right)^2+S_0\, , \ena where $S_0$ can be set
to zero as argued above. So, for the Ricci flat horizon case, the
entropy is proportional to the horizon area and the log term
disappears. The difference from the well-known area formula in
Einstein's general relativity is the additional factor $(1+s)/(2-s)$
or $\sqrt{(3\lambda-1)/2}$ in the black hole entropy. The free
energy is given by
\begin{equation}
F_{e}=\gamma(s-1)M.
\end{equation}
The global thermodynamic stability is discussed in Sec.~3.

\subsection{Non-Einstein Case: $\lambda\rightarrow 1/3$}

In the case, notice the definition of the speed of light in
(\ref{eq5}),  we see that the temperature of the black hole is still
well defined. However, the mass and entropy of the black hole
diverge. To have a finite result in this case, one could take a
rescaling of the speed of light so that $(c^3/G)\sqrt{3\lambda-1}$
goes to a ${\rm finite\ constant}$ in the limit $\lambda \to 1/3$.
But the physical meaning (if any) is not clear at the moment for
this rescaling.

\section{Local Stability of Black Hole Thermodynamics}

By studying Euclidean action, or free energy, we can find
information on the global stability of black hole thermodynamics.
However, to discuss the local stability, we have to calculate the
heat capacity of black holes. From the expressions for the mass and
temperature, we get heat capacity as
\begin{equation}
C=\frac{\partial M}{\partial T}= \frac{c^3 }{4G}
\left(\Omega_k\ell^2\right)\cdot\left(\frac{1+s}{2-s}\right)\cdot
\left\{
\frac{\left(k+r_+^2/\ell^2\right)\left[(2-s)r_+^2/\ell^2-ks\right]}{ks^2+
(s-1)(s-2)r_+^2/\ell^2}\right\}\, ,
\end{equation}
For the case $\lambda=1$ or $s=1/2$, this result goes back to the
one in~\cite{CCO}. Here, we give some discussions on
thermodynamical stability of the black holes.
\begin{enumerate}
\item $k=0$:
It is easy to find, for $k=0$, the dominator will
change sign at $s=1$. So for $s> 1$ case, the Ricci flat black holes
are thermodynamically unstable. Note that in this case, the black
hole is also globally unstable according to its free energy. This
behavior of the heat capacity for this case can be directly found from
Fig.~\ref{HC_0}.
\begin{figure}[htb]
\centering
\begin{minipage}[c]{.58\textwidth}
\centering
\includegraphics[width=\textwidth]{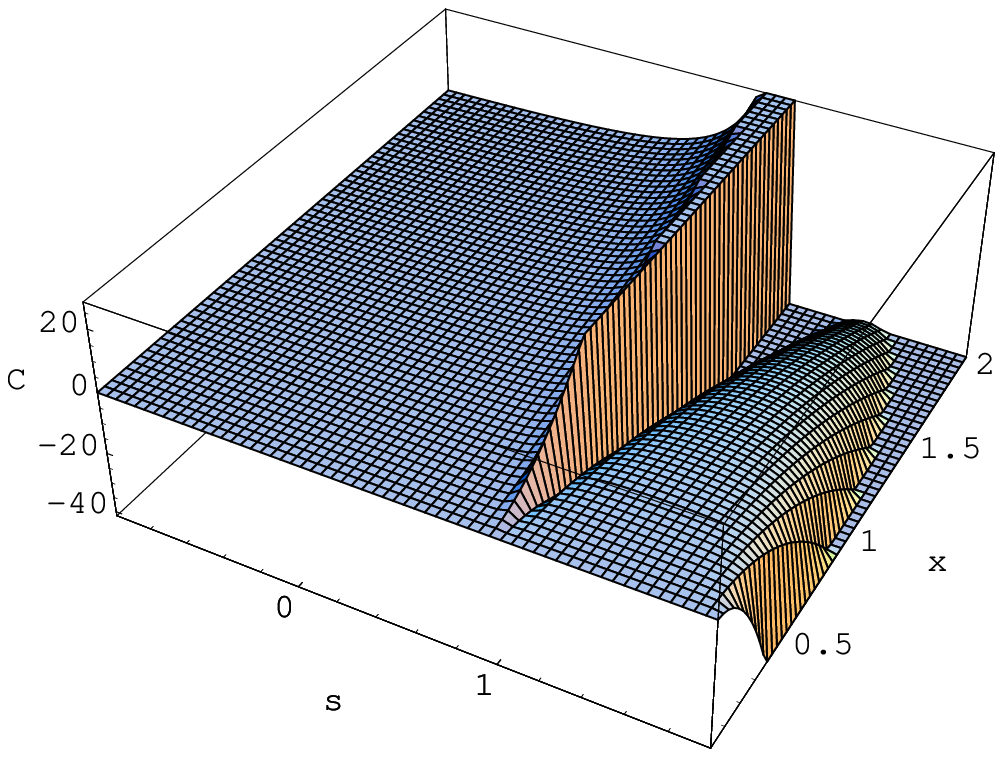}
\caption{Heat capacity for $k=0$, where $x=r_+/\ell$. } \label{HC_0}
\end{minipage}
\end{figure}

\item $k=-1$:
In this case, the two factors in the numerator of the heat capacity
are both positive: the first one is positive because the minimal
horizon is at $r_+=\ell$ for massless black hole; the second comes
from the requirement of the positive definiteness of Hawking
temperature. Thus the sign of the heat capacity is completely
determined by the denominator. Therefore when $s \ge 1$, the heat
capacity is always negative. When $-1< s<1$, the heat capacity is
positive for $r_+^2/\ell^2>s^2/(1-s)(2-s)$. Otherwise, it is
negative and it diverges when $r_+^2/\ell^2=s^2/(1-s)(2-s)$. Note
that requiring $r_+^2/\ell^2 \ge 1$ leads to $s \ge 2/3$, which
means that for $ 2/3 <s <1$, the black hole is thermodynamically
stable if $r_+^2/\ell^2>s^2/(1-s)(2-s)$. Fig.~\ref{HC_-1} depicts
the heat capacity for $r_{+}>\ell$.

\begin{figure}[htb]
\centering
\begin{minipage}[c]{.58\textwidth}
\centering
\includegraphics[width=\textwidth]{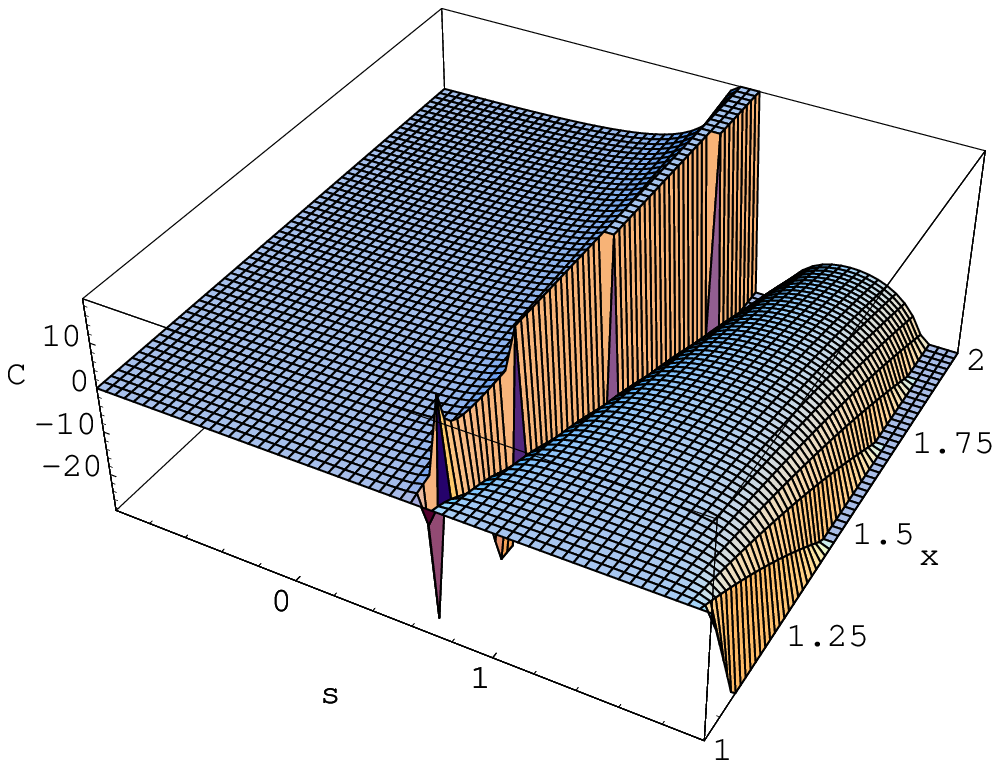}
\caption{Heat capacity for $k=-1$, where $x=r_+/\ell$. }
\label{HC_-1}
\end{minipage}
\end{figure}


\item  $k=1$: In this case, the positive definiteness of
temperature demands $r_+^2/\ell^2 \ge s/(2-s)$. When the equality
holds, it corresponds to an extremal black hole with vanishing
temperature. Then we can see that when $-1 <s \le 1$, the heat
capacity is always positive. When $ 1 < s <2$, it is positive for
\begin{equation}
\sqrt{\frac{s}{2-s}}<r_+/\ell<\sqrt{\frac{s^2}{(s-1)(2-s)}},
\end{equation}
and  it becomes negative  for
$r_+/\ell>\sqrt{\frac{s^2}{(s-1)(2-s)}}$ . One can find this
behavior in Fig.~\ref{HC_1}. It is interesting to note that in this
case even for a large black hole, it is not thermodynamically
stable.

\begin{figure}[htb]
\centering
\begin{minipage}[c]{.58\textwidth}
\centering
\includegraphics[width=\textwidth]{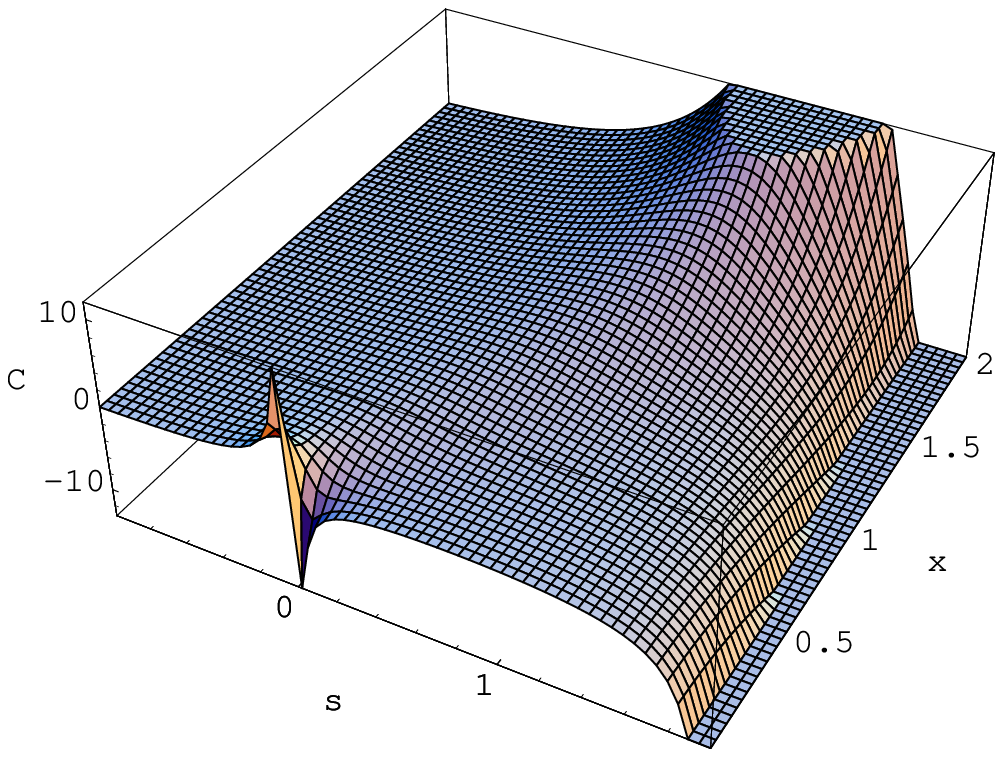}
\caption{Heat capacity for $k=1$, where $x=r_+/\ell$. } \label{HC_1}
\end{minipage}
\end{figure}

\end{enumerate}

In summary we have found that in three cases with different horizon
topologies, there always exist locally thermodynamically stable phases and
unstable phases in suitable parameter regimes.

\section{Conclusion and Discussions}

In this paper, using the canonical Hamiltonian method we have
generalized the discussion of thermodynamics of topological black
holes for $\lambda=1$ case~\cite{CCO} to the general $\lambda\ge
1/3$ case. All the thermodynamical quantities we have got reduce
to those in~\cite{CCO} when $\lambda=1$, although we have used
different approaches. We have also studied the global and local
thermodynamical stability of these black holes, and found that
there exist rich phase structures compared to the case of AdS
Schwarzschild black holes in Einstein's general relativity. In all
three different horizon topologies, locally stable or unstable
phases exist in the proper parameter space. It is quite different
from the case in Einstein's general relativity.



For general $\lambda$, up to a constant, the entropy of  black
hole not only receives a logarithm correction, but also gets a multiplicative
factor which is a function of $\lambda$ or $s$. This function reduces to one
when $\lambda=1$. In this case, the one quarter in the relation between
the entropy and the horizon area is recovered, the well-known entropy formula
in units of $c=G=1$.
This means that the area formula of the entropy is not restored in the
infrared region of the theory unless the dynamical coupling constant is one.

\section*{Acknowledgments}
This work was supported partially by grants from NSFC, China (No.
10821504 and No. 10525060), a grant from the Chinese Academy of
Sciences with No.KJCX3-SYW-N2, the Grant-in-Aid for Scientific
Research Fund of the JSPS No. 20540283, and the Japan-U.K. Research
Cooperative Program, the YST program of Asia Pacific Center for
Theoretical Physics, Korea.


\begin{thebibliography}{99}
\bibitem{Horava}
  P.~Horava,
  Phys.\ Rev.\  D {\bf 79} (2009) 084008
  [arXiv:0901.3775 [hep-th]];
  JHEP {\bf 0903} (2009) 020
  [arXiv:0812.4287 [hep-th]];
  arXiv:0902.3657 [hep-th].

\bibitem{Visser}
  M.~Visser,
  arXiv:0902.0590 [hep-th].
\bibitem{Maccione}
  L.~Maccione, A.~M.~Taylor, D.~M.~Mattingly and S.~Liberati,
  JCAP {\bf 0904} (2009) 022
  [arXiv:0902.1756 [astro-ph.HE]].
\bibitem{Carvalho}
  P.~R.~S.~Carvalho and M.~M.~Leite,
  arXiv:0902.1972 [hep-th].
\bibitem{Volovich}
  A.~Volovich and C.~Wen,
  arXiv:0903.2455 [hep-th].
\bibitem{Jenkins}
  A.~Jenkins,
  arXiv:0904.0453 [gr-qc].
\bibitem{Takahashi}
  T.~Takahashi and J.~Soda,
  arXiv:0904.0554 [hep-th].
\bibitem{Calcagni}
  G.~Calcagni,
  arXiv:0904.0829 [hep-th].
\bibitem{Kiritsis}
  E.~Kiritsis and G.~Kofinas,
  arXiv:0904.1334 [hep-th].
\bibitem{Kluson}
  J.~Kluson,
  arXiv:0904.1343 [hep-th].
\bibitem{LMP}
  H.~Lu, J.~Mei and C.~N.~Pope,
  arXiv:0904.1595 [hep-th].
\bibitem{Mukohyama}
  S.~Mukohyama,
  arXiv:0904.2190 [hep-th].
\bibitem{Brandenberger}
  R.~Brandenberger,
  arXiv:0904.2835 [hep-th].
\bibitem{Nikolic}
  H.~Nikolic,
  arXiv:0904.3412 [hep-th].
\bibitem{Nastase}
  H.~Nastase,
  arXiv:0904.3604 [hep-th].
\bibitem{CCO}
  R.~G.~Cai, L.~M.~Cao and N.~Ohta,
  arXiv:0904.3670 [hep-th].
\bibitem{Cai}
  R.~G.~Cai, Y.~Liu and Y.~W.~Sun,
  arXiv:0904.4104 [hep-th].
\bibitem{Piao}
  Y.~S.~Piao,
  arXiv:0904.4117 [hep-th].
\bibitem{Gao1}
  X.~Gao,
  arXiv:0904.4187 [hep-th].
\bibitem{Colgain}
  E.~O.~Colgain and H.~Yavartanoo,
  arXiv:0904.4357 [hep-th].
\bibitem{Chen}
  B.~Chen and Q.~G.~Huang,
  arXiv:0904.4565 [hep-th].
\bibitem{Mukohyama1}
  S.~Mukohyama, K.~Nakayama, F.~Takahashi and S.~Yokoyama,
  arXiv:0905.0055 [hep-th].

\bibitem{MK}
  Y.~S.~Myung and Y.~W.~Kim,
  arXiv:0905.0179 [hep-th].

\bibitem{Cai:2009dx}
  R.~G.~Cai, B.~Hu and H.~B.~Zhang,
  arXiv:0905.0255 [hep-th].

\bibitem{Orlando}
  D.~Orlando and S.~Reffert,
  arXiv:0905.0301 [hep-th].

\bibitem{Gao2}
  C.~Gao,
  arXiv:0905.0310 [astro-ph.CO].


\bibitem{Nishioka:2009iq}
  T.~Nishioka,
  arXiv:0905.0473 [hep-th].

\bibitem{Kehagias:2009is}
  A.~Kehagias and K.~Sfetsos,
  arXiv:0905.0477 [hep-th].












\bibitem{BTZ}M.~Banados, C.~Teitelboim and J.~Zanelli,
  Phys.\ Rev.\  D {\bf 49}, 975 (1994)
  [arXiv:gr-qc/9307033].

\bibitem{CS}R.~G.~Cai and K.~S.~Soh,
  Phys.\ Rev.\  D {\bf 59}, 044013 (1999)
  [arXiv:gr-qc/9808067].

\bibitem{RT}T.~Regge and C.~Teitelboim,
  Annals Phys.\  {\bf 88}, 286 (1974).

\bibitem{MTZ}
  C.~Martinez, R.~Troncoso and J.~Zanelli,
  Phys.\ Rev.\  D {\bf 70} (2004) 084035
  [arXiv:hep-th/0406111].


\bibitem{Peca:1998dv}
  C.~S.~Peca and J.~P.~S.~Lemos,
  J.\ Math.\ Phys.\  {\bf 41}, 4783 (2000)
  [arXiv:gr-qc/9809029].


\bibitem{Peca:1998cs}
  C.~S.~Peca and J.~P.~S.~Lemos,
  Phys.\ Rev.\  D {\bf 59}, 124007 (1999)
  [arXiv:gr-qc/9805004].

\bibitem{Ashtekar:1984}
Ashtekar A and Magnon A 1984, Asymptotically anti-de Sitter space-times, Class.
Quantum Grav. 1 L39

\bibitem{Ashtekar:1999jx}
  A.~Ashtekar and S.~Das,
  Class.\ Quant.\ Grav.\  {\bf 17}, L17 (2000)
  [arXiv:hep-th/9911230].


\end{thebibliography}
\end{document}